\documentclass[usenatbib]{mn2e}
\usepackage{journals}
\usepackage{graphicx}
\usepackage{upgreek}
\usepackage{amssymb}

\newcommand{\ergs}{\rm\,erg\,s^{-1}}

\newcommand{\sax}{SAX\,J1808.4$-$3658}
\newcommand{\psr}{PSR\,J1023+0038}
\newcommand{\igr}{PSR\,J1824$-$2452I}
\newcommand{\xss}{XSS\,J12270$-$4859}
\newcommand{\MSun}{$\rm{M}_{\odot}$}

\title[A state change in the low-mass X-ray binary
  XSS\,J12270$-$4859]{A state change in the low-mass X-ray binary
  XSS\,J12270$-$4859}

\author[Bassa et al.]  {C.\,G.\,Bassa$^1$\thanks{email: bassa@astron.nl}, 
  A.\,Patruno$^{2,1}$, 
  J.\,W.\,T.\,Hessels$^{1,3}$, 
  E.\,F.\,Keane$^{4,5}$, 
  B.\,Monard$^6$,\newauthor
  E.\,K.\,Mahony$^1$,
  S.\,Bogdanov$^7$,
  S.\,Corbel$^8$,
  P.\,G.\,Edwards$^9$,
  A.\,M.\,Archibald$^1$,\newauthor
  G.\,H.\,Janssen$^1$,
  B.\,W.\,Stappers$^{10}$,
  S.\,Tendulkar$^{11}$\\ 
  $^1$ASTRON, the Netherlands Institute for Radio Astronomy, Postbus 2, 7990 AA, Dwingeloo, the Netherlands\\ 
  $^2$Leiden Observatory, Leiden University, Postbus 9513, 2300 RA, Leiden, the Netherlands\\
  $^3$Astronomical Institute 'Anton Pannekoek', University of Amsterdam, Postbus 94249, 1090 GE, Amsterdam, the Netherlands\\
  $^4$Centre for Astrophysics and Supercomputing, Swinburne University of Technology, Mail H30, PO Box 218, VIC 3122, Australia\\
  $^5$ARC Centre of Excellence for All-Sky Astrophysics (CAASTRO)\\
  $^6$Kleinkaroo Observatory, Center for Backyard Astrophysics Kleinkaroo,
Sint Helena 1B, PO Box 281, Calitzdorp 6660, South Africa\\
  $^7$Columbia Astrophysics Laboratory, Columbia University, 550 West 120th Street, New York, NY 10027, USA\\
  $^8$CEA, CNRS, Universit\'e Paris Diderot, Sorbonne Paris Cit\'e, AIM, UMR 7158, DSM, IRFU, SAp, F-91191 Gif sur Yvette, France\\
  $^9$CSIRO Astronomy and Space Science, Australia Telescope National Facility, P.O. Box 76, Epping, NSW 1710, Australia\\
  $^{10}$Jodrell Bank Centre for Astrophysics, School of Physics and Astronomy, The University of Manchester, Manchester M13 9PL, UK\\
  $^{11}$Space Radiation Laboratory, California Institute of Technology, 1200 E California Blvd, MC 249-17, Pasadena, CA 91125, USA\\
}

\begin{document}

\date{Accepted \today. Received \today; in original form \today}

\pagerange{\pageref{firstpage}--\pageref{lastpage}} \pubyear{2002}

\maketitle

\label{firstpage}

\begin{abstract}
Millisecond radio pulsars acquire their rapid rotation rates through
mass and angular momentum transfer in a low-mass X-ray binary
system. Recent studies of \igr\ and \psr\ have observationally
demonstrated this link, and they have also shown that such systems can
repeatedly transition back-and-forth between the radio millisecond
pulsar and low-mass X-ray binary states. This also suggests that a
fraction of such systems are not newly born radio millisecond pulsars
but are rather suspended in a back-and-forth state switching phase,
perhaps for giga-years. \xss\ has been previously suggested to be a
low-mass X-ray binary, and until recently the only such system to be
seen at MeV-GeV energies. We present radio, optical and X-ray
observations that offer compelling evidence that \xss\ is a low-mass
X-ray binary which transitioned to a radio millisecond pulsar state
between 2012 November 14 and 2012 December 21.  Though radio
pulsations remain to be detected, we use optical and X-ray
photometry/spectroscopy to show that the system has undergone a sudden
dimming and no longer shows evidence for an accretion disk. The
optical observations constrain the orbital period to
$6.913\pm0.002$\,hr.
\end{abstract}


\section{Introduction}

Low-mass X-ray binaries (LMXBs) are systems in which a neutron star or
black hole is orbited by a Roche-lobe-filling, main-sequence companion
with a mass $\lesssim 1$\,\MSun.  In systems with neutron star
primaries the accretion disk can transfer both matter and angular
momentum to the neutron star, thereby spinning it up and eventually
producing a ``recycled" millisecond radio pulsar (MSP;
\citealt{acrs82,rs82}).  The LMXB-MSP evolutionary link has been
demonstrated by, e.g., (i) the observation of accretion-powered
pulsations in \sax\ \citep{wk98}; (ii) the discovery of the MSP \psr,
a system in which an accretion disk was previously observed
\citep{asr+09}; and (iii) the transition of \igr/IGR\,J18245$-$245
from an MSP to an accreting X-ray millisecond pulsar (AMXP) and back
\citep{pfb+13}.  More recently, \psr\ has re-entered a radio-quiet,
accretion-disk state which shows much of the same X-ray and optical
phenomenology observed in \igr\ \citep{sah+13,tll+13,pah+14} ---
though the system has not yet entered a state of full accretion onto
the neutron star surface.  As such, we have witnessed a growing number
of systems that straddle the classical definitions of LMXB and MSP,
and which are helping us map this interesting evolutionary phase.

In addition to the aforementioned systems, recent radio pulsar
searches have identified dozens of new ``black widow" and ``redback"
systems \citep{rob13}.  These are compact radio pulsar binary systems
in which eclipsing of the pulsed signal is normally observed around
superior conjunction of the neutron star, indicating that material is
actively being ablated from the companion by the pulsar wind. The
black widow systems are those in which the companion is very low mass
($\sim 0.01$\,\MSun) and, likely degenerate.  The redbacks have likely
non-degenerate companions that are significantly more massive (about
0.1 to 0.7\,\MSun).  The rapid growth in the known population of such
sources has come from deep, targeted radio pulsation searches towards
Galactic globular clusters (e.g.\ \citealt{rhs+05,hrs+07}) as well as
towards unidentified $\gamma$-ray sources found with the {\it Fermi
  Gamma-ray Space Telescope} \citep{rap+12}.  Since \sax, \psr\ and
\igr\ can all be classified as belonging to the black widow or redback
families (though \sax\ has never been seen to pulse in radio it has
shown evidence that a radio pulsar is active during the X-ray
quiescent phase, see \citealt{bsa+03} and \citealt{cac+04}), it seems
likely that in the coming decade some of these other, newly found MSPs
will also be seen to transition to an X-ray active phase and back.  In
fact, one may ponder whether these systems are gradually making a
definitive transition from LMXB to MSP, or whether they are stuck
transitioning back and forth for many giga-years (see \citealt{ccth13}
for evolutionary modeling of such systems).

\psr's recent transition back to an LMXB-like state --- in which an
accretion disk has returned \citep{pah+14} --- has been accompanied by
an additional observational surprise: namely that the MeV-GeV
brightness of the system has quintupled in concert with the
disappearance of the radio pulsar \citep{sah+13}. The emission of
$\gamma$-rays is somewhat unexpected in LMXBs, and is more commonly
seen in X-ray binaries with O or B-type companions in much wider
orbits \citep{dub13}. This is the first time that $\gamma$-ray
variability has been observed in an LMXB/MSP system, and it provides a
new diagnostic tool while also raising new questions about the origin
of this high-energy emission.

Prior to the increase of the \psr\ system's $\gamma$-ray flux, the
only LMXB thought to emit $\gamma$-rays was \xss. Discovered in the
RXTE slew survey \citep{sr04}, this system was initially classified as
an $R=15.7$ cataclysmic variable by \citet{mmp+06}, based on the
presence of optical emission lines. Follow-up optical and X-ray
observations cast doubt on this classification, suggesting instead
that it is an LMXB \citep{pre09,stei09,mfb+10}. Subsequently,
\citet{mfb+10} and \citet{hsc+11} noted the positional coincidence of
\xss\ with the \textit{Fermi} $\gamma$-ray source 1FGL\,J1227.9$-$4852
(2FGL\,J1227.7$-$4853). Based on that coincidence and similarities
with \psr, \citet{hsc+11}, \citet{mbf+13} and \citet{ptl13} suggested
that also \xss\ could harbour an active radio MSP.

Here we present radio, optical and X-ray observations to expand on our
earlier report \citep{bph+13} where we suggested that \xss\ has
recently undergone a state transition similar to that observed in
\psr.  In the case of \xss, however, it appears that the system has
transitioned from an LMXB-like state to a radio MSP-like state,
i.e.\ the opposite switch that \psr\ has recently made \citep{sah+13}.
In \S\ref{Obs} we present our observations and analysis, in
\S\ref{Res} our results, and end with a discussion in \S\ref{Disc}.

\section{Observations and analysis}
\label{Obs}

\subsection{Optical}
\begin{figure}
  \includegraphics[angle=270,width=8cm]{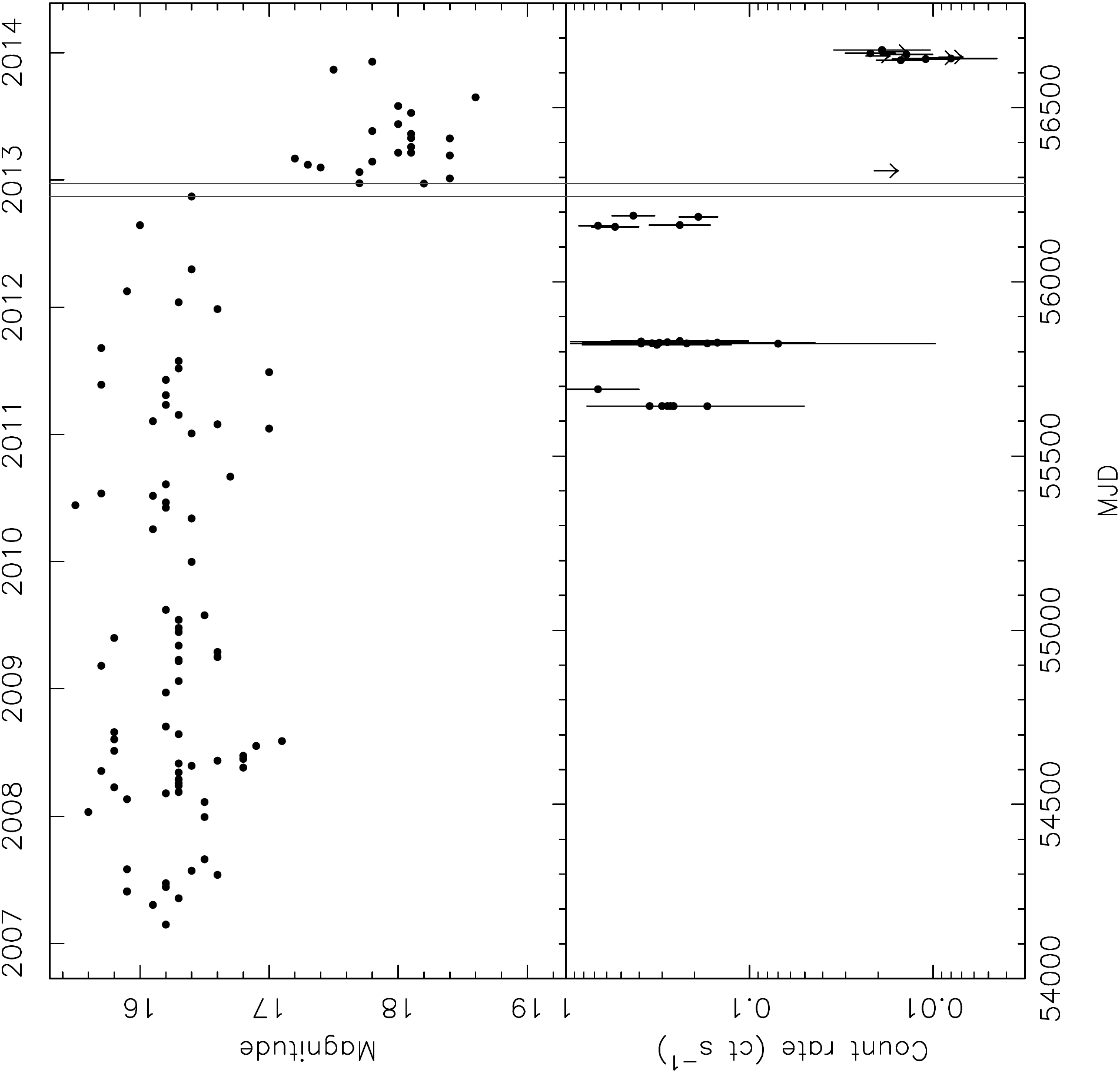}
  \caption{\textit{Top panel:} The long-term optical light-curve of
    \xss\ obtained from the Bronberg and Kleinkaroo
    Observatories. Unfiltered magnitudes were determined from images
    obtained with the same telescope and CCD camera combination over
    the 6-year period. Typical magnitude uncertainties range from
    0.02\,mag at 16th magnitude to 0.25\,mag at 19th magnitude.  A 1.5
    to 2\,mag decrease in brightness occurred between 2012 November 14
    and 2012 December 21. That time period is indicated with the
    vertical lines. \textit{Bottom panel:} \textit{Swift}/XRT
    long-term light-curve of \xss\ in the 0.3-10\,keV energy band. The
    plot shows an order-of-magnitude decrease in the X-ray count rate
    that is qualitatively consistent with the decrease in optical
    brightness. The data prior to 2012 are taken from \citet{mbf+13}.}
  \label{fig:lc}
\end{figure}

XSS\,J12270$-$4859 has been monitored at optical wavelengths with a
roughly monthly cadence or better since 2007. Observations were
carried out at the Bronberg (2007-2010) and Kleinkaroo Observatories
(2011-2013) in South Africa. Unfiltered magnitudes were obtained with
Meade RCX 400 telescopes, having apertures of 30 and 35\,cm and
working at F/8, and an SBIG ST8-XME CCD camera, using $2\times2$
spatial binning and averages of 3 to 5 exposures of 13\,s each. The
CCD camera has a quantum efficiency of 50\% to 90\% over the 4500 to
8000\,\AA\ wavelength range, roughly covering the Johnson $V$ and
$R$-band filters. The light-curve is shown in Fig.\,\ref{fig:lc}.

\begin{figure*}
  \includegraphics[width=\textwidth]{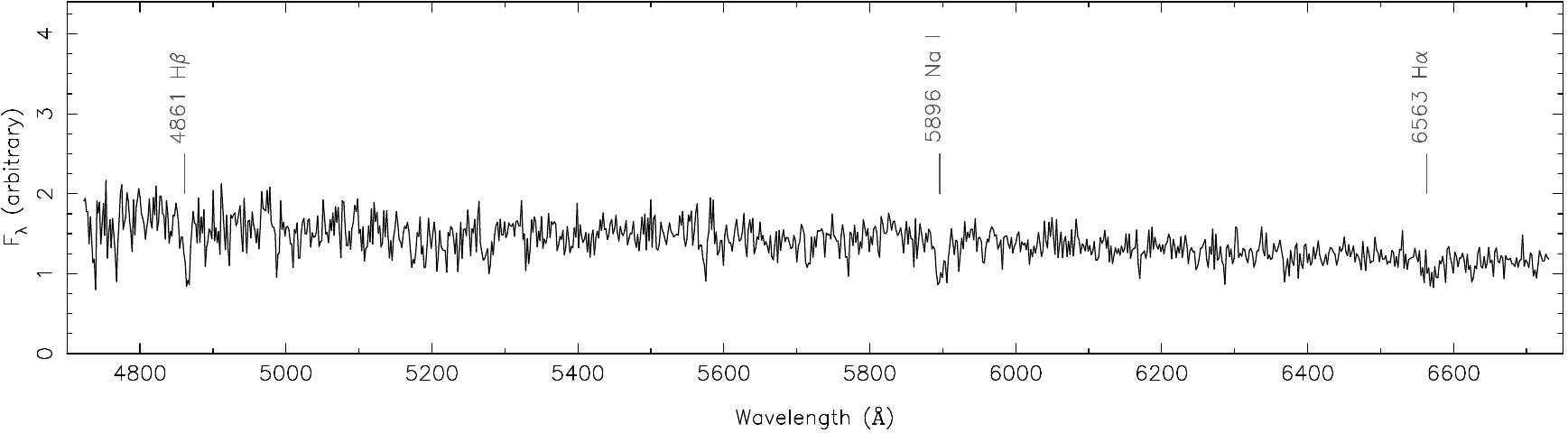}
  \caption{The average optical spectrum of XSS\,J12270$-$4859 obtained
    with EFOSC2 at the NTT on 2013 November 8 (orbital phase of
    $\phi=0.66$). Obvious spectral lines are indicated. The spectral
    type is consistent with late-G/early-K.}
  \label{fig:spectra}
\end{figure*}

We obtained optical spectroscopy of \xss\ with EFOSC2, the ESO Faint
Object Spectrograph and Camera, at the NTT on La Silla in Chile on
2013 November 8.  The sky was clear with $1\farcs0$ seeing. Two 900-s
exposures were obtained with a $1\arcsec$ slit in combination with
grism \#18, giving a wavelength coverage of 4720 to 6730\,\AA. With
$2\times2$ binning, this is sampled at 2.0\,\AA\,pix$^{-1}$ and
provides a resolution of 8.4\,\AA. The observations were corrected for
bias offsets and the sky background was subtracted using clean regions
offset from the spectral trace. The spectra were extracted using the
optimal extraction method of \citet{hor86} and wavelength calibrated
using arc lamp exposures. Flux calibration was performed against the
Feige\,110 spectrophotometric flux standards. The calibration is
approximate as no attempt was made to correct for slit
losses. Figure\,\ref{fig:spectra} shows the average of the two 900\,s
spectra.

These ground-based optical data were complemented by eleven pointed
observations made with the \textit{Swift}/Ultra-Violet and Optical
Telescope (UVOT; \citealt{rkm+05}).  The first observation was taken
on 2013 January 26, whereas the next ten were collected between 2013
December 10 and 2014 January 8. Each of the latter ten observations
consists of one to three exposures, all taken with the $U$-band filter
(central wavelength of 3465\,\AA). The 2013 January observation
instead comprises six short snapshots recorded with each of the six
optical and UV filters.  A visual inspection of the images reveals a
dim counterpart in all the $U$-band filter observations and a lack of
counterpart in five of the six short snapshots of January 26 (only the
$B$-band filter counterpart is clearly seen). 

We extracted the source magnitude using the standard UVOT pipeline and
selected the photons coming from a region of $3-4\arcsec$ centered on
the pixel with the highest count rate (and compatible, within
uncertainties, with the known source position of \citealt{mmp+06}).
The background is extracted from a circular region with radius of
$20\arcsec$, positioned in a location free from bright sources.  We
extracted the magnitudes with two methods: in the first method, we
summed all the single exposures comprised in each observation with the
\texttt{uvotimsum} tool to obtain the highest possible signal-to-noise
ratio and extracted the magnitudes with the \texttt{uvotsource} tool
(distributed with the \textit{heasoft} v.6.14). The second method uses the tool
\texttt{uvotmaghist} to extract the magnitudes from each single
exposure and follow the source brightness variations on a timescale of
hours.

We also make use of recently acquired data using the
\textit{XMM-Newton} Optical Monitor (OM) data; the corresponding EPIC
X-ray data set will be presented elsewhere. \textit{XMM-Newton}
observed \xss\ as a target of opportunity starting on 2013 December 29
for a total duration of 38 ks. The OM was configured in fast mode to
permit rapid photometry using the $U$-band filter. Ten exposures of
3\,ks were acquired. The data were processed using the {\tt omfchain}
pipeline in SAS\footnote{The \textit{XMM-Newton} SAS is developed and
  maintained by the Science Operations Centre at the European Space
  Astronomy Centre and the Survey Science Centre at the University of
  Leicester.} version xmmsas\_20130501\_1901-13.0.0 using the default
parameters.

\subsection{X-ray}

The same eleven \textit{Swift} observations, plus five additional
ones, were used to extract the X-ray data recorded with the X-Ray
Telescope (XRT).  The additional observations refer to the period from
2012 August to September. Each of the 17 observations lasted for
$0.5-2.5$\,ks and the XRT operated exclusively in PC mode with a time
resolution of approximately 2.5\,s. A number of additional
\textit{Swift}/XRT observations were collected between 2005 and 2011;
their analysis was previously reported in \cite{mbf+13} and
\cite{tkl13}.

The data were analyzed using the XRT pipeline and by applying standard
screening criteria. We extracted all photons with an energy between
$0.3-10$\,keV that fall within an extraction region of
$15-40\arcsec$. The center of the extraction region is centered on the
pixel with the highest count rate and is compatible with the best
astrometric position available \citep{mmp+06}.  We then extracted the
background by selecting a region with a radius two times the source
extraction region size, randomly placed but in locations far from
known X-ray sources. The count rate was corrected for the presence of
bad pixels, vignetting and dead columns, and normalized to match the
source region area.  We then repeated the entire procedure by using
the \textit{detect} and \textit{sosta}
FTOOLS\footnote{\texttt{http://heasarc.gsfc.nasa.gov/ftools/}} under
the \textit{ximage} package (v.4.5.1) and obtained consistent
results. For the non-detections, 95\% upper limits are given,
calculated according to the prescription given in \citet{geh86}. The
small number of photons prevented the fit of the X-ray spectrum but
the count-rate is consistently below, at least by an
order-of-magnitude, that reported in \citet{mfb+10,mbf+13} in all
eleven observations recorded after 2012 December.

\subsection{Radio}
A faint, continuum radio source has previously been associated with
\xss\ \citep{hsc+11}. In order to probe its variability, we conducted
a new radio observation with the Australia Compact Array Telescope
(ATCA) located near Narrabri (Australia), on 2013, December 17. The
observations were conducted at 5.5 and 9\,GHz in the 750B array
configuration with the upgraded and sensitive CABB back-end
\citep{wfa+11} for a total time on source of 5.48\,hours. The
amplitude and band-pass calibrator was PKS\,1934$-$638, and the
antennas' gain and phase calibration, as well as the polarization
leakage, were derived from regular observations of the nearby
calibrator PMN\,J1326$-$5256. The editing, calibration, Fourier
transformation with multifrequency algorithms, deconvolution, and
image analysis were performed using the {\tt MIRIAD} software package
\citep{sau98}.

Furthermore, we searched for radio pulsations using the 64-m Parkes
radio telescope.  Observations were acquired at 1.4\,GHz using the
central beam of the multi-beam receiver (pointing position:
$\alpha_\mathrm{J2000} = 12^\mathrm{h}27^\mathrm{m}58\fs68$,
$\delta_\mathrm{J2000} = -48\degr53^{\prime}42\farcs0$).  Summed
polarization, filterbank data were recorded as 2-bit samples over a
400-MHz bandwidth, of which 340\,MHz is usable, using the BPSR backend
\citep{kjs+10}, which provided 0.39-MHz channels and 64-$\upmu$s time
resolution.  \xss\ was observed for 5\,hrs on 2013 November 13 and for
1\,hr on 2013 November 17.  

Using the
PRESTO\footnote{http://www.cv.nrao.edu/$\sim$sransom/presto/} pulsar
search suite, we excised radio frequency interference (RFI) and
performed an acceleration search \citep{rem02} for spectral drifts
between $-500 < z < 500$ bins, as well as trial dispersion measures
(DMs) between $0-300$\,pc\,cm$^{-3}$ (in steps of
$0.1$\,pc\,cm$^{-3}$).  The highest trial DM was chosen to be
$2\times$ larger than the maximum expected DM along this
line-of-sight, according to the NE2001 model of the Galactic free
electron density \citep{cl02}. If \xss\ is at 1.4--3.6\,kpc, as
estimated by \citet{mbf+13}, then the model predicts a DM in the range
of 30 to 100\,pc\,cm$^{-3}$.  For $\mathrm{DM} = 100$\,pc\,cm$^{-3}$,
the intra-channel dispersion smearing at the lowest observed frequency
is 200\,$\upmu$s.  Acceleration search processing gives improved
sensitivity to periodic signals that are Doppler shifted by orbital
motion, but it assumes a linear drift of the signal in the power
spectrum and thus is only valid when the observation time,
$T_\mathrm{obs}$, is $\lesssim 0.1 P_\mathrm{orb}$, the orbital period
\citep{rem02}.  Given the proposed 6.913-hr orbital period (see
\S\ref{Res}), we searched the data in 10 and 32-minute chunks, in
order to remain in the constant acceleration regime.  The resulting
candidates were sifted to look for signals that showed a peak in
signal-to-noise ratio with DM.  Promising candidates were folded at
the candidate rotational period and DM, in order to create a full
diagnostic plot.  The likelihood that the resulting signals were
astrophysical in origin was then judged based on a range of standard
criteria \citep{hrs+07}.
 
\begin{figure}
  \includegraphics[width=8cm]{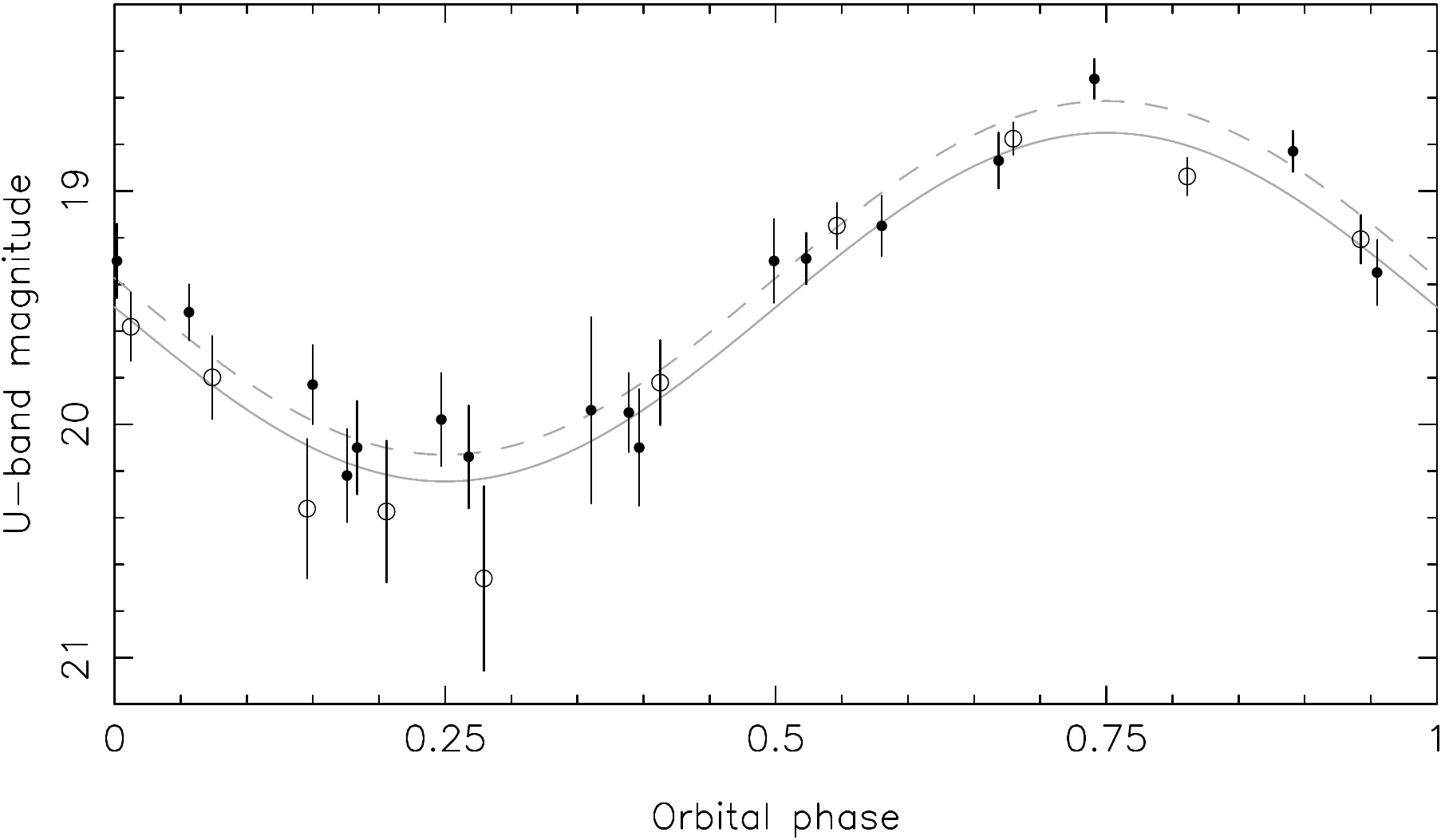}
  \caption{The $U$-band magnitudes observed by \textit{Swift}/UVOT
    from 2013 December 10 to 2014 January 8 (solid points) and by
    \textit{XMM-Newton}/OM on 2013 December 29/30 (open points). The
    data are folded at the best fit orbital period, and the dashed and
    solid lines show the best fit amplitude and mean for
    \textit{Swift}/UVOT and \textit{XMM-Newton}/OM, respectively. The
    curves are offset since the data of each instrument was allowed to
    have a different mean and amplitude while fitting for the same
    orbital period and phase. }
  \label{fig:u-band_lc}
\end{figure}

\section{Results}
\label{Res}

The optical and X-ray observations presented here allow us to study
the behaviour of \xss\ on long as well as short time
scales. Figure\,\ref{fig:lc} shows the long-term optical and X-ray
light-curves of \xss. Both light-curves show a sudden decrease in
brightness, where the decrease appears to have occurred between 2012
November 14 and 2012 December 21. A decrease of 1.5 to 2\,mag is seen
in the optical, while the X-ray light-curve shows that the
\textit{Swift}/XRT count rate decreased by at least a factor of 10
over that same time period.

On shorter time scales, the 10-hr $U$-band light-curve obtained by
\textit{XMM-Newton}/OM shows sinusoidal variations at the 6.91-hr
orbital period reported by \citet{mcm+13}. The variability is also
present in the $U$-band observations obtained with \textit{Swift}/UVOT
over the 29-day interval. By approximating the light-curves of both
instruments with a sine and fitting for six parameters: period, phase,
and, to allow for differences between the $U$-band filters, mean
$U$-band magnitude and variability amplitude separately for each
instrument, we obtain an orbital period of $6.913\pm0.002$\,hr, with
zero phase occurring at HJD $2456651.026\pm0.002$. The $\chi^2$ of the
fit is 19.2 for 21 degrees of freedom. Figure\,\ref{fig:u-band_lc}
shows the $U$-band observations folded on this ephemeris. This orbital
period is consistent, though an order-of-magnitude more accurate, with
that reported by \citet{mcm+13}.

The optical spectrum of \xss\ obtained in 2013 November is shown in
Fig.\,\ref{fig:spectra} and shows weak absorption lines of
Hydrogen, as well as the sodium doublet. The spectrum does not have
the signal-to-noise, nor covers a large enough spectral range, to
determine accurate spectral parameters, but a comparison with template
spectra from the library of \citet{bbp+03} constrains the spectral
type to late-G/early-K.

To estimate the X-ray luminosity we assume an absorbed power-law with
a very low absorption column ($N_{H}{\sim}10^{21}\rm\,cm^{-2}$),
compatible with that estimated by \citet{mfb+10}, and spectral index
$\Gamma = 1-2$. Given the source distance (1.4--3.6\,kpc;
\citealt{mbf+13}), this corresponds to an unabsorbed luminosity
between $7\times10^{31}\ergs$ and $4\times10^{32}\ergs$. With the
correspondingly low count rates, we cannot explore timescales shorter
than the length of each observation (${\sim}1\,$ks). Furthermore, the
small number of detections and low count rate are insufficient to
study X-ray orbital modulation.

The 2013 December ATCA observations lead to no detection of the radio
source that was previously found \citep{hsc+11} at the location of
\xss\ with $3\sigma$ upper limits of 30\,$\upmu$Jy at 5.5\,GHz and
33\,$\upmu$Jy at 9\,GHz. A reanalysis of the 2010 radio flux densities
presented in \citet{hsc+11}, taking into account the primary beam
correction, yields detections of $190\pm30$\,$\upmu$Jy at 5.5\,GHz and
$180\pm40$\,$\upmu$Jy at 9\,GHz. This clearly indicates a significant
reduction (by at least a factor of 6) of radio emission from \xss\ in
association with its recent state change.

The Parkes searches have thus far failed to identify a radio pulsar
signal.  In our 32-minute searches, we can infer a flux density upper
limit of 0.2\,mJy at 1.4\,GHz, assuming a minimum detectable
signal-to-noise S/N$_{\rm min} = 10$ and a 20\% duty cycle.  This is
comparable to the flux density limit reported during \xss's previous
state \citep{hsc+11}.  The radio non-detection is discussed in more
detail in \S\ref{Disc}.

\section{Discussion and conclusions}
\label{Disc}

Our radio, optical, and X-ray observations show that, sometime between
2012 November 14 and 2012 December 21, \xss\ transitioned to a new
state where it is consistently and considerably fainter in these
bands, and where the previous signs of an accretion disk
(i.e.\ double-peaked optical emission lines) have
disappeared. Subsequent to our report of this transition in
\citet{bph+13}, \citet{tkl13} reported a similar decrease in
brightness of \xss\ in $\gamma$-rays, while \citet{cmm+14} confirm the
absence of emission lines at all orbital phases.

This phenomenology shows many parallels with \psr, which has recently
transitioned from being a radio millisecond pulsar to a state where it
resembles a quiescent LMXB
\citep{asr+09,wat+09,sah+13,tll+13,pah+14}. Compared with \psr's
recent behaviour, however, the \xss\ observations presented here
strongly suggest the reverse transition, i.e.\ from an LMXB-like state
in 2012 to one where the accretion disk is absent in 2013. Besides the
presence of an accretion disk during the LMXB state, the \xss\ X-ray
light-curve presented by \citet{mbf+13} displays frequent low flux
states in brightness, similar to what is seen in the recent LMXB state
of \psr\ \citep{pah+14} as well as during \igr's X-ray active phase
\citep{pfb+13}. The cause of the low X-ray flux states in both systems
is presently unknown. The sinusoidal $U$-band light-curve that is seen
in the present state of \xss\ is comparable to the light-curves
presented in \citet{wwp04} and \citet{ta05} and can be understood as
irradiation of the companion by the neutron star.

A radio source coincident with \xss\ was detected by \citet{hsc+11} in
ATCA observations obtained in 2009 when the source was in the LMXB
state. Our follow-up ATCA observations obtained after the transition
show it has decreased in brightness by at least a factor of 6. Recent
Very Large Array observations of \psr\ after it returned to the LMXB
state show a flat spectrum radio source (Deller et al.\ in prep.). All
these parallels between \xss\ and \psr\ lead us to conclude that
\xss\ harbours an active rotation-powered millisecond pulsar (see also
\citealt{ptl13}) and, based on the present information, one could
classify \xss\ as a ``redback'' system in its current state.

Thus far, however, we have failed to detect radio pulsations from
\xss.  Though such a detection would show without a doubt that the
system has transitioned from an LMXB state to an MSP state, the lack
of radio detection in no way rules out such a state transition.  Given
the high inferred mass ratio, $q=0.53$ \citep{mcm+13}, it appears that
\xss\ would be observed as a redback in its present state. In analogy
with other redback systems, we expect the pulsar to be eclipsed for
$\sim 50$\% of the time at $\sim 1.4$\,GHz, and it is possible that
the system is enshrouded a much larger fraction of the time.  Indeed,
the recent examples of PSR\,J2339$-$0533 \citep{rs11} and
PSR\,J1311$-$3430 \citep{rrc+13} show that the radio detectability of
such pulsars may be exceedingly poor, even though the radio pulsar
mechanism is clearly active. For this reason, continued radio pulsar
searches may eventually be successful in detecting the source. It is
also simply possible that the radio pulsar is too weak to detect with
the current Parkes data, too highly accelerated without using a far
more complex acceleration search, or not beamed towards the Earth.

\psr's transition from MSP to LMXB was accompanied by a 5-fold
increase in $\gamma$-ray flux \citep{sah+13}. Assuming that \xss\ is
mirroring such a transition, we would thus expect a marked decrease in
$\gamma$-ray brightness. Using the optically and X-ray-derived epoch
of the state change, \citet{tkl13} report that the $\gamma$-ray
brightness has indeed decreased by a factor of 1.5 to 2, but note that
the decrease appears more gradual than what was seen in \psr. Indeed,
the $\gamma$-ray light-curve of \xss\ appears more complicated than
that of \psr\, and we are investigating whether there is contamination
from another source.

\psr's transition from radio MSP to LMXB is constrained to have
happened within a 2-week window \citep{sah+13}, and may well have
happened even more abruptly.  \xss's mirror transition from LMXB to an
MSP-like state is constrained to a 5-week period, and is thus also
quite rapid.  An additional comparison is provided by \igr's 2013
transition from accreting X-ray MSP to an observable radio MSP, within
a period of less than 3 weeks.  It thus appears that the
back-and-forth transitioning of such systems between MSP and LMXB is a
rapid process which high-cadence radio and X-ray monitoring can
constrain further.  In particular, though practically difficult to
achieve, near-daily, joint radio and X-ray monitoring would either
detect or strongly constrain whether there is a lag between the X-ray
brightening and disappearance of the radio MSP.

Besides the similarities with \psr, there are also differences. \xss's
high reported mass ratio $q = 0.53$ \citep{mcm+13} suggests a
significantly more massive companion compared with
\psr\ ($M_\mathrm{c}=0.2$\,\MSun) and the majority of the other
observed redbacks (see \citealt{rob13} for a review). Such a high
companion mass is not unprecedented, however; PSR\,J1723$-$2837
\citep{cls+13} has an inferred $M_\mathrm{c,min} =
0.7$\,\MSun\ companion and PSR\,J2129$-$0428 (Hessels et al., in
prep.) has a $M_\mathrm{c,min} = 0.5$\,\MSun\ companion. A much larger
number of LMXB/MSP transition systems will have to be characterized
before it is possible to determine the role that companion mass has on
the frequency of state switches.

Thus far we have observed similar back-and-forth transitioning from
three redbacks; \psr, \igr, and now also in \xss.  Should we expect
the same from the black widows?  For the accreting X-ray millisecond
pulsars, roughly half have black-widow-like companions and half have
redback-like companions \citep{pw12}. There is no obvious correlation
between the recurrence timescales in such system and the companion
mass.  Why have such systems not been observed as radio pulsars?
Possibly simply because they are mostly very distant, which is
problematic both because radio MSPs are intrinsically quite faint and
because distant sources at low Galactic latitudes will be scattered.

The observations and analysis presented in this paper show that
\xss\ has undergone a state transition similar to those observed in
\psr\ and \igr. The analysis of recently obtained \textit{XMM/Newton}
and \textit{Chandra} X-ray observations of \xss\ may shed further
light on the presence of an active radio MSP, though multiwavelength
observations are certainly warranted. For example, the present absence
of the accretion disk allows for optical spectroscopy of the main
sequence companion. This will provide constraints on the masses of the
companion and the neutron star. Finally, regular optical and X-ray
will help identify future transitions, constraining the timescales
involved in the transitions and possibly trigger detailed
investigations during the transitions.

\section*{Acknowledgments}
C.G.B.\ acknowledges support from ERC Advanced Grant ``LEAP'' (227947,
PI: Michael Kramer). A.M.A.\ and J.W.T.H.\ acknowledge support from a
Vrije Competitie grant from NWO.  J.W.T.H.\ and A.P.\ acknowledge
support from NWO Vidi grants. J.W.T.H.\ also acknowledges funding from
an ERC Starting Grant ``DRAGNET'' (337062). E.F.K.\ acknowledges the
support of the Australian Research Council Centre of Excellence for
All-sky Astrophysics (CAASTRO), through project number
CE110001020. S.C.\ acknowledges the financial support from the
UnivEarthS Labex program of Sorbonne Paris Cit\'e (ANR-10-LABX-0023
and ANR-11-IDEX-0005-02).  The Australia Telescope Compact Array and
Parkes radio telescope are part of the Australia Telescope National
Facility which is funded by the Commonwealth of Australia for
operation as a National Facility managed by CSIRO.  The work presented
was based in part on observations obtained with \textit{XMM-Newton},
an ESA science mission with instruments and contributions directly
funded by ESA Member States and NASA.

\bibliographystyle{mn}

\begin{thebibliography}{44}
\expandafter\ifx\csname natexlab\endcsname\relax\def\natexlab#1{#1}\fi

\bibitem[{{Alpar} {et~al.}(1982){Alpar}, {Cheng}, {Ruderman}, \&
  {Shaham}}]{acrs82}
{Alpar} M.~A., {Cheng} A.~F., {Ruderman} M.~A., {Shaham} J., 1982, \nat, 300,
  728

\bibitem[{{Archibald} {et~al.}(2009){Archibald}, {Stairs}, {Ransom}, {Kaspi},
  {Kondratiev}, {Lorimer}, {McLaughlin}, {Boyles}, {Hessels}, {Lynch}, {van
  Leeuwen}, {Roberts}, {Jenet}, {Champion}, {Rosen}, {Barlow}, {Dunlap}, \&
  {Remillard}}]{asr+09}
{Archibald} A.~M., {Stairs} I.~H., {Ransom} S.~M., {Kaspi} V.~M., {Kondratiev}
  V.~I., {Lorimer} D.~R., {McLaughlin} M.~A., {Boyles} J., {Hessels} J.~W.~T.,
  {Lynch} R., {van Leeuwen} J., {Roberts} M.~S.~E., {Jenet} F., {Champion}
  D.~J., {Rosen} R., {Barlow} B.~N., {Dunlap} B.~H., {Remillard} R.~A., 2009,
  Science, 324, 1411

\bibitem[{{Bassa} {et~al.}(2013){Bassa}, {Patruno}, {Hessels}, {Archibald},
  {Mahony}, {Monard}, {Keane}, {Bogdanov}, {Stappers}, {Janssen}, \&
  {Tendulkar}}]{bph+13}
{Bassa} C.~G., {Patruno} A., {Hessels} J.~W.~T., {Archibald} A.~M., {Mahony}
  E.~K., {Monard} B., {Keane} E.~F., {Bogdanov} S., {Stappers} B.~W., {Janssen}
  G.~H., {Tendulkar} S., 2013, The Astronomer's Telegram, 5647, 1

\bibitem[{{Burderi} {et~al.}(2003){Burderi}, {Di Salvo}, {D'Antona}, {Robba},
  \& {Testa}}]{bsa+03}
{Burderi} L., {Di Salvo} T., {D'Antona} F., {Robba} N.~R., {Testa} V., 2003,
  \aap, 404, L43

\bibitem[{{Campana} {et~al.}(2004){Campana}, {D'Avanzo}, {Casares}, {Covino},
  {Israel}, {Marconi}, {Hynes}, {Charles}, \& {Stella}}]{cac+04}
{Campana} S., {D'Avanzo} P., {Casares} J., {Covino} S., {Israel} G., {Marconi}
  G., {Hynes} R., {Charles} P., {Stella} L., 2004, \apjl, 614, L49

\bibitem[{{Casares Velazquez} {et~al.}(2014){Casares Velazquez}, {de Martino},
  {Mason}, {D'Avanzo}, {Campana}, {Fugazza}, {Covino}, {Belloni},
  {Munoz-Darias}, {Cornelisse}, \& {Nicastro}}]{cmm+14}
{Casares Velazquez} J., {de Martino} D., {Mason} E., {D'Avanzo} P., {Campana}
  S., {Fugazza} S., {Covino} S., {Belloni} T., {Munoz-Darias} T., {Cornelisse}
  R., {Nicastro} L., 2014, The Astronomer's Telegram, 5747, 1

\bibitem[{{Chen} {et~al.}(2013){Chen}, {Chen}, {Tauris}, \& {Han}}]{ccth13}
{Chen} H.-L., {Chen} X., {Tauris} T.~M., {Han} Z., 2013, \apj, 775, 27

\bibitem[{{Cordes} \& {Lazio}(2002)}]{cl02}
{Cordes} J.~M., {Lazio} T.~J.~W., 2002, \texttt{astro-ph/0207156}

\bibitem[{{Crawford} {et~al.}(2013){Crawford}, {Lyne}, {Stairs}, {Kaplan},
  {McLaughlin}, {Freire}, {Burgay}, {Camilo}, {D'Amico}, {Faulkner}, {Kramer},
  {Lorimer}, {Manchester}, {Possenti}, \& {Steeghs}}]{cls+13}
{Crawford} F., {Lyne} A.~G., {Stairs} I.~H., {Kaplan} D.~L., {McLaughlin}
  M.~A., {Freire} P.~C.~C., {Burgay} M., {Camilo} F., {D'Amico} N., {Faulkner}
  A., {Kramer} M., {Lorimer} D.~R., {Manchester} R.~N., {Possenti} A.,
  {Steeghs} D., 2013, \apj, 776, 20

\bibitem[{{de Martino} {et~al.}(2013{\natexlab{a}}){de Martino}, {Belloni},
  {Falanga}, {Papitto}, {Motta}, {Pellizzoni}, {Evangelista}, {Piano},
  {Masetti}, {Bonnet-Bidaud}, {Mouchet}, {Mukai}, \& {Possenti}}]{mbf+13}
{de Martino} D., {Belloni} T., {Falanga} M., {Papitto} A., {Motta} S.,
  {Pellizzoni} A., {Evangelista} Y., {Piano} G., {Masetti} N., {Bonnet-Bidaud}
  J.-M., {Mouchet} M., {Mukai} K., {Possenti} A., 2013{\natexlab{a}}, \aap,
  550, A89

\bibitem[{{de Martino} {et~al.}(2013{\natexlab{b}}){de Martino}, {Casares
  Velazquez}, {Mason}, {Kotze}, \& {Buckley}}]{mcm+13}
{de Martino} D., {Casares Velazquez} J., {Mason} E., {Kotze} M., {Buckley}
  D.~A.~H., 2013{\natexlab{b}}, The Astronomer's Telegram, 5651, 1

\bibitem[{{de Martino} {et~al.}(2010){de Martino}, {Falanga}, {Bonnet-Bidaud},
  {Belloni}, {Mouchet}, {Masetti}, {Andruchow}, {Cellone}, {Mukai}, \&
  {Matt}}]{mfb+10}
{de Martino} D., {Falanga} M., {Bonnet-Bidaud} J.-M., {Belloni} T., {Mouchet}
  M., {Masetti} N., {Andruchow} I., {Cellone} S.~A., {Mukai} K., {Matt} G.,
  2010, \aap, 515, A25

\bibitem[{{Dubus}(2013)}]{dub13}
{Dubus} G., 2013, \aapr, 21, 64

\bibitem[{{Gehrels}(1986)}]{geh86}
{Gehrels} N., 1986, \apj, 303, 336

\bibitem[{{Hessels} {et~al.}(2007){Hessels}, {Ransom}, {Stairs}, {Kaspi}, \&
  {Freire}}]{hrs+07}
{Hessels} J.~W.~T., {Ransom} S.~M., {Stairs} I.~H., {Kaspi} V.~M., {Freire}
  P.~C.~C., 2007, \apj, 670, 363

\bibitem[{{Hill} {et~al.}(2011){Hill}, {Szostek}, {Corbel}, {Camilo}, {Corbet},
  {Dubois}, {Dubus}, {Edwards}, {Ferrara}, {Kerr}, {Koerding}, {Kozie{\l}}, \&
  {Stawarz}}]{hsc+11}
{Hill} A.~B., {Szostek} A., {Corbel} S., {Camilo} F., {Corbet} R.~H.~D.,
  {Dubois} R., {Dubus} G., {Edwards} P.~G., {Ferrara} E.~C., {Kerr} M.,
  {Koerding} E., {Kozie{\l}} D., {Stawarz} {\L}., 2011, \mnras, 415, 235

\bibitem[{{Horne}(1986)}]{hor86}
{Horne} K., 1986, \pasp, 98, 609

\bibitem[{{Keith} {et~al.}(2010){Keith}, {Jameson}, {van Straten}, {Bailes},
  {Johnston}, {Kramer}, {Possenti}, {Bates}, {Bhat}, {Burgay}, {Burke-Spolaor},
  {D'Amico}, {Levin}, {McMahon}, {Milia}, \& {Stappers}}]{kjs+10}
{Keith} M.~J., {Jameson} A., {van Straten} W., {Bailes} M., {Johnston} S.,
  {Kramer} M., {Possenti} A., {Bates} S.~D., {Bhat} N.~D.~R., {Burgay} M.,
  {Burke-Spolaor} S., {D'Amico} N., {Levin} L., {McMahon} P.~L., {Milia} S.,
  {Stappers} B.~W., 2010, \mnras, 409, 619

\bibitem[{{Le Borgne} {et~al.}(2003){Le Borgne}, {Bruzual}, {Pell{\' o}},
  {Lan{\c c}on}, {Rocca-Volmerange}, {Sanahuja}, {Schaerer}, {Soubiran}, \&
  {V{\'{\i}}lchez-G{\' o}mez}}]{bbp+03}
{Le Borgne} J.-F., {Bruzual} G., {Pell{\' o}} R., {Lan{\c c}on} A.,
  {Rocca-Volmerange} B., {Sanahuja} B., {Schaerer} D., {Soubiran} C.,
  {V{\'{\i}}lchez-G{\' o}mez} R., 2003, \aap, 402, 433

\bibitem[{{Masetti} {et~al.}(2006){Masetti}, {Morelli}, {Palazzi}, {Galaz},
  {Bassani}, {Bazzano}, {Bird}, {Dean}, {Israel}, {Landi}, {Malizia},
  {Minniti}, {Schiavone}, {Stephen}, {Ubertini}, \& {Walter}}]{mmp+06}
{Masetti} N., {Morelli} L., {Palazzi} E., {Galaz} G., {Bassani} L., {Bazzano}
  A., {Bird} A.~J., {Dean} A.~J., {Israel} G.~L., {Landi} R., {Malizia} A.,
  {Minniti} D., {Schiavone} F., {Stephen} J.~B., {Ubertini} P., {Walter} R.,
  2006, \aap, 459, 21

\bibitem[{{Papitto} {et~al.}(2013{\natexlab{a}}){Papitto}, {Ferrigno}, {Bozzo},
  {Rea}, {Pavan}, {Burderi}, {Burgay}, {Campana}, {di Salvo}, {Falanga},
  {Filipovi{\'c}}, {Freire}, {Hessels}, {Possenti}, {Ransom}, {Riggio},
  {Romano}, {Sarkissian}, {Stairs}, {Stella}, {Torres}, {Wieringa}, \&
  {Wong}}]{pfb+13}
{Papitto} A., {Ferrigno} C., {Bozzo} E., {Rea} N., {Pavan} L., {Burderi} L.,
  {Burgay} M., {Campana} S., {di Salvo} T., {Falanga} M., {Filipovi{\'c}}
  M.~D., {Freire} P.~C.~C., {Hessels} J.~W.~T., {Possenti} A., {Ransom} S.~M.,
  {Riggio} A., {Romano} P., {Sarkissian} J.~M., {Stairs} I.~H., {Stella} L.,
  {Torres} D.~F., {Wieringa} M.~H., {Wong} G.~F., 2013{\natexlab{a}}, \nat,
  501, 517

\bibitem[{{Papitto} {et~al.}(2013{\natexlab{b}}){Papitto}, {Torres}, \&
  {Li}}]{ptl13}
{Papitto} A., {Torres} D.~F., {Li} J., 2013{\natexlab{b}}, \mnras

\bibitem[{{Patruno} {et~al.}(2014){Patruno}, {Archibald}, {Hessels},
  {Bogdanov}, {Stappers}, {Bassa}, {Janssen}, {Kaspi}, {Tendulkar}, \&
  {Lyne}}]{pah+14}
{Patruno} A., {Archibald} A.~M., {Hessels} J.~W.~T., {Bogdanov} S., {Stappers}
  B.~W., {Bassa} C.~G., {Janssen} G.~H., {Kaspi} V.~M., {Tendulkar} S., {Lyne}
  A.~G., 2014, \apjl, 781, L3

\bibitem[{{Patruno} \& {Watts}(2012)}]{pw12}
{Patruno} A., {Watts} A.~L., 2012, ArXiv e-prints, 1206.2727

\bibitem[{{Pretorius}(2009)}]{pre09}
{Pretorius} M.~L., 2009, \mnras, 395, 386

\bibitem[{{Radhakrishnan} \& {Srinivasan}(1982)}]{rs82}
{Radhakrishnan} V., {Srinivasan} G., 1982, Current Science, 51, 1096

\bibitem[{{Ransom} {et~al.}(2002){Ransom}, {Eikenberry}, \&
  {Middleditch}}]{rem02}
{Ransom} S.~M., {Eikenberry} S.~S., {Middleditch} J., 2002, \aj, 124, 1788

\bibitem[{{Ransom} {et~al.}(2005){Ransom}, {Hessels}, {Stairs}, {Freire},
  {Camilo}, {Kaspi}, \& {Kaplan}}]{rhs+05}
{Ransom} S.~M., {Hessels} J.~W.~T., {Stairs} I.~H., {Freire} P.~C.~C., {Camilo}
  F., {Kaspi} V.~M., {Kaplan} D.~L., 2005, Science, 307, 892

\bibitem[{{Ray} {et~al.}(2012){Ray}, {Abdo}, {Parent}, {Bhattacharya},
  {Bhattacharyya}, {Camilo}, {Cognard}, {Theureau}, {Ferrara}, {Harding},
  {Thompson}, {Freire}, {Guillemot}, {Gupta}, {Roy}, {Hessels}, {Johnston},
  {Keith}, {Shannon}, {Kerr}, {Michelson}, {Romani}, {Kramer}, {McLaughlin},
  {Ransom}, {Roberts}, {Saz Parkinson}, {Ziegler}, {Smith}, {Stappers},
  {Weltevrede}, \& {Wood}}]{rap+12}
{Ray} P.~S., {Abdo} A.~A., {Parent} D., {Bhattacharya} D., {Bhattacharyya} B.,
  {Camilo} F., {Cognard} I., {Theureau} G., {Ferrara} E.~C., {Harding} A.~K.,
  {Thompson} D.~J., {Freire} P.~C.~C., {Guillemot} L., {Gupta} Y., {Roy} J.,
  {Hessels} J.~W.~T., {Johnston} S., {Keith} M., {Shannon} R., {Kerr} M.,
  {Michelson} P.~F., {Romani} R.~W., {Kramer} M., {McLaughlin} M.~A., {Ransom}
  S.~M., {Roberts} M.~S.~E., {Saz Parkinson} P.~M., {Ziegler} M., {Smith}
  D.~A., {Stappers} B.~W., {Weltevrede} P., {Wood} K.~S., 2012, ArXiv e-prints, 1205.3089

\bibitem[{{Ray} {et~al.}(2013){Ray}, {Ransom}, {Cheung}, {Giroletti},
  {Cognard}, {Camilo}, {Bhattacharyya}, {Roy}, {Romani}, {Ferrara},
  {Guillemot}, {Johnston}, {Keith}, {Kerr}, {Kramer}, {Pletsch}, {Saz
  Parkinson}, \& {Wood}}]{rrc+13}
{Ray} P.~S., {Ransom} S.~M., {Cheung} C.~C., {Giroletti} M., {Cognard} I.,
  {Camilo} F., {Bhattacharyya} B., {Roy} J., {Romani} R.~W., {Ferrara} E.~C.,
  {Guillemot} L., {Johnston} S., {Keith} M., {Kerr} M., {Kramer} M., {Pletsch}
  H.~J., {Saz Parkinson} P.~M., {Wood} K.~S., 2013, \apjl, 763, L13

\bibitem[{{Roberts}(2013)}]{rob13}
{Roberts} M.~S.~E., 2013, in IAU Symposium, Vol. 291, IAU Symposium, pp.
  127--132

\bibitem[{{Romani} \& {Shaw}(2011)}]{rs11}
{Romani} R.~W., {Shaw} M.~S., 2011, \apjl, 743, L26

\bibitem[{{Roming} {et~al.}(2005){Roming}, {Kennedy}, {Mason}, {Nousek}, {Ahr},
  {Bingham}, {Broos}, {Carter}, {Hancock}, {Huckle}, {Hunsberger}, {Kawakami},
  {Killough}, {Koch}, {McLelland}, {Smith}, {Smith}, {Soto}, {Boyd},
  {Breeveld}, {Holland}, {Ivanushkina}, {Pryzby}, {Still}, \& {Stock}}]{rkm+05}
{Roming} P.~W.~A., {Kennedy} T.~E., {Mason} K.~O., {Nousek} J.~A., {Ahr} L.,
  {Bingham} R.~E., {Broos} P.~S., {Carter} M.~J., {Hancock} B.~K., {Huckle}
  H.~E., {Hunsberger} S.~D., {Kawakami} H., {Killough} R., {Koch} T.~S.,
  {McLelland} M.~K., {Smith} K., {Smith} P.~J., {Soto} J.~C., {Boyd} P.~T.,
  {Breeveld} A.~A., {Holland} S.~T., {Ivanushkina} M., {Pryzby} M.~S., {Still}
  M.~D., {Stock} J., 2005, \ssr, 120, 95

\bibitem[{{Saitou} {et~al.}(2009){Saitou}, {Tsujimoto}, {Ebisawa}, \&
  {Ishida}}]{stei09}
{Saitou} K., {Tsujimoto} M., {Ebisawa} K., {Ishida} M., 2009, \pasj, 61, L13

\bibitem[{{Sault} \& {Killeen}(1998)}]{sau98}
{Sault} R.~J., {Killeen} N.~E.~B., 1998, The Miriad User's Guide. Sydney:
  Australia Telescope National Facility

\bibitem[{{Sazonov} \& {Revnivtsev}(2004)}]{sr04}
{Sazonov} S.~Y., {Revnivtsev} M.~G., 2004, \aap, 423, 469

\bibitem[{{Stappers} {et~al.}(2013){Stappers}, {Archibald}, {Hessels}, {Bassa},
  {Bogdanov}, {Janssen}, {Kaspi}, {Lyne}, {Patruno}, {Tendulkar}, {Hill}, \&
  {Glanzman}}]{sah+13}
{Stappers} B.~W., {Archibald} A.~M., {Hessels} J.~W.~T., {Bassa} C.~G.,
  {Bogdanov} S., {Janssen} G.~H., {Kaspi} V.~M., {Lyne} A.~G., {Patruno} A.,
  {Tendulkar} S., {Hill} A.~B., {Glanzman} T., 2013, ArXiv e-prints, 1311.7506

\bibitem[{{Takata} {et~al.}(2013){Takata}, {Li}, {Leung}, {Kong}, {Tam}, {Hui},
  {Wu}, {Xing}, {Cao}, {Tang}, {Wang}, \& {Cheng}}]{tll+13}
{Takata} J., {Li} K.~L., {Leung} G.~C.~K., {Kong} A.~K.~H., {Tam} P.~H.~T.,
  {Hui} C.~Y., {Wu} E.~M.~H., {Xing} Y., {Cao} Y., {Tang} S., {Wang} Z.,
  {Cheng} K.~S., 2013, ArXiv e-prints, 1312.0605

\bibitem[{{Tam} {et~al.}(2013){Tam}, {Kong}, \& {Li}}]{tkl13}
{Tam} P.~H.~T., {Kong} A.~K.~H., {Li} K.~L., 2013, The Astronomer's Telegram,
  5652, 1

\bibitem[{{Thorstensen} \& {Armstrong}(2005)}]{ta05}
{Thorstensen} J.~R., {Armstrong} E., 2005, \aj, 130, 759

\bibitem[{{Wang} {et~al.}(2009){Wang}, {Archibald}, {Thorstensen}, {Kaspi},
  {Lorimer}, {Stairs}, \& {Ransom}}]{wat+09}
{Wang} Z., {Archibald} A.~M., {Thorstensen} J.~R., {Kaspi} V.~M., {Lorimer}
  D.~R., {Stairs} I., {Ransom} S.~M., 2009, \apj, 703, 2017

\bibitem[{{Wijnands} \& {van der Klis}(1998)}]{wk98}
{Wijnands} R., {van der Klis} M., 1998, \nat, 394, 344

\bibitem[{{Wilson} {et~al.}(2011){Wilson}, {Ferris}, {Axtens}, {Brown},
  {Davis}, {Hampson}, {Leach}, {Roberts}, {Saunders}, {Koribalski}, {Caswell},
  {Lenc}, {Stevens}, {Voronkov}, {Wieringa}, {Brooks}, {Edwards}, {Ekers},
  {Emonts}, {Hindson}, {Johnston}, {Maddison}, {Mahony}, {Malu}, {Massardi},
  {Mao}, {McConnell}, {Norris}, {Schnitzeler}, {Subrahmanyan}, {Urquhart},
  {Thompson}, \& {Wark}}]{wfa+11}
{Wilson} W.~E., {Ferris} R.~H., {Axtens} P., {Brown} A., {Davis} E., {Hampson}
  G., {Leach} M., {Roberts} P., {Saunders} S., {Koribalski} B.~S., {Caswell}
  J.~L., {Lenc} E., {Stevens} J., {Voronkov} M.~A., {Wieringa} M.~H., {Brooks}
  K., {Edwards} P.~G., {Ekers} R.~D., {Emonts} B., {Hindson} L., {Johnston} S.,
  {Maddison} S.~T., {Mahony} E.~K., {Malu} S.~S., {Massardi} M., {Mao} M.~Y.,
  {McConnell} D., {Norris} R.~P., {Schnitzeler} D., {Subrahmanyan} R.,
  {Urquhart} J.~S., {Thompson} M.~A., {Wark} R.~M., 2011, \mnras, 416, 832

\bibitem[{{Woudt} {et~al.}(2004){Woudt}, {Warner}, \& {Pretorius}}]{wwp04}
{Woudt} P.~A., {Warner} B., {Pretorius} M.~L., 2004, \mnras, 351, 1015

\end{thebibliography}

\label{lastpage}

\end{document}